\documentclass[aps,prd,twocolumn,nofootinbib]{revtex4-1}
\usepackage{graphicx}
\usepackage{amsmath}
\usepackage{bm}
\usepackage{slashed}
\usepackage{epsfig}
\usepackage{amsfonts}
\usepackage{epstopdf}
\usepackage{color}
\usepackage{extarrows}
\usepackage{multirow}
\usepackage[utf8]{inputenc}
\usepackage{hyperref}

\def\be{\begin{equation}}
\def\ee{\end{equation}}
\def\ba{\begin{eqnarray}}
\def\ea{\end{eqnarray}}
 \def\la{ \langle}
  \def\ra{ \rangle}
     
      \def\r{ \gamma}

\allowdisplaybreaks

\begin{document}

\title{The $\rho$-meson electromagnetic form factors within the light-front quark model}

\author{Shuai Xu$^{a}$}
\email{xushuai@zknu.edu.cn}
\author{Xiao-Nan Li$^{b}$}
\email{lixn@tlu.edu.cn (Corresponding author)}
\author{Xing-Gang Wu$^{c}$}
\email{wuxg@cqu.edu.cn}

\affiliation{
$^a$School of Physics and Telecommunications Engineering, Zhoukou Normal University, Henan 466001, P.R. China\\
$^b$School of Electrical Engineering, Tongling University, Anhui 244000, P.R. China\\
$^c$Department of Physics, Chongqing Key Laboratory for Strongly Coupled Physics, Chongqing University, Chongqing 401331, P.R. China}

\date{\today}

\begin{abstract}
	
In this paper, we study the $\rho$-meson electromagnetic form factors (EMFFs) within the framework of light-front quark model (LFQM). The physical form factors $G_{C,M,Q}(Q^2)$ of $\rho$-meson as well as the charged square radius $\langle r^2\rangle$, the magnetic moment $\mu$ and the quadrupole moment $\bar Q$ are calculated, which describe the behaviors of EMFFs at zero momentum transfer. Using the type-II replacement, we find that the zero-mode does contribute zero to the matrix element $S_{00}^+$. It is found that the ``$M\to M_{0}$" replacement improves angular condition remarkably, which permits different prescriptions of $\rho$-meson EMFFs give the consistent results. The residue tiny violation of angular condition needs other explanations than the zero-mode contributions. Our results indicate that the relativistic effects or interaction internal structure are weaken in the zero-binding limit. This work is also applied for the other spin-1 particles.

\end{abstract}

\maketitle

\section{Introduction}

The electromagnetic form factors (EMFFs) include essential information on the hadronic structure and interaction between the constituents, which are important and fundamental ingredients in phenomenal researches~\cite{Brodsky:1992px, Chung:1991st, Chung:1988mu, Chung:1989zz, deMelo:1997hh, deMelo:1997cb, Jaus:1999zv, Jaus:2002sv, Choi:1997iq, Cao:1998jm, Huang:2004fn, Choi:2004ww}. In the past few years, a lot of efforts to study electromagnetic structure of the spin-1 particles in light-front approaches have been performed. The works~\cite{Grach:1983hd, Chung:1988my, Frankfurt:1988gn} have researched the EMFFs of the deuteron, in which the deuteron is treated as the two-nucleon system. The universal properties of electromagnetic interactions in spin-1 bound states are investigated in Ref.~\cite{Brodsky:1992px}. The authors of Refs.~\cite{Bakker:2002mt,Choi:2004ww,deMelo:1997cb,Jaus:2002sv} have examined extensively the zero-mode effects in EMFFs of $\rho$ meson in several aspects.

Light-front quark model provides a conceptually simple, feasible framework for the determination of the quantities involved the non-perturbative information, such as the kinds of form factors, decay constants and distribution amplitudes~\cite{Jaus:1999zv, Jaus:1989au, Jaus:1996np, Cheng:2003sm, Choi:2013mda, Choi:2017uos, Chang:2018zjq, Chang:2019mmh}. In practice, these quantities are usually extracted through the definitions in the special matric element of current operator and the computational task are mainly focused on the determination of matrix element, correspondingly. The current operator in matrix element is usually approximated by a one-body current, which means the initial and final states have the same number of constituents~\cite{Brodsky:1998hn}. Without containing interaction dependent parts (two-body current), which is necessarily included in the angular momentum operator, this leads to the lost of rotational covariance of the light-front matric element~\cite{Jaus:2002sv}. Consequently, those approaches do not permit an unambiguous determination of the physical quantities for some cases, such as the decay constants of vector mesons and weak transition form factors between vector mesons. Previous works~\cite{Brodsky:1992px, Grach:1983hd, Chung:1988my} show that different descriptions of deuteron EMFFs by using different extraction methods do not give a consistent result, this is called as ``the self-consistency" problem. As proposed in Ref.~\cite{Grach:1983hd}, the break of rotational covariance brings about untenable angular condition, which is a key relationship between matrix elements in case of EMFFs, and the violation of angular condition will undoubtedly bring about the above non-uniqueness issue. The angular condition could be restored at this level of approximation only if the zero-mode contributions are reasonably considered~\cite{Brodsky:1998hn, deMelo:1998an, deMelo:1999gn, Choi:1998nf, Jaus:2002sv, Yan:1973qg}. Once the angular condition is hold, different prescriptions of EMFFs will achieve same predictions. Refs.\cite{Brodsky:1998hn, deMelo:1998an, deMelo:1999gn, Choi:1998nf} suggest that the zero-mode contributions are naturally associated with the covariance of light-front matrix element. Refs.\cite{Jaus:1999zv, Jaus:2002sv} develop a systematic approach to calculate the zero-mode contributions, which also permits a consistent determination of the physical form factors.

A promising remedy called ``type-II" replacement has been suggested to deal with the vector meson decay constant $f_V$~\cite{Choi:2013mda}, which avoids the self-consistency problem. Inspired by this treatment, Refs.\cite{Chang:2018zjq, Chang:2019mmh} perform the calculation of the meson decay constants and the weak transition form factors, in which the self-consistency problem and the non-covariant problem have also been carefully discussed. Furthermore, it has been pointed out that the self-consistency problem is closely associated with the covariant problem, and those two issues have the same origin that the ambiguous decomposition for the terms related with the fixed lightlike vector $\omega$. Interestingly, numerical analysis show that even though the zero-mode exist formally, their contributions will be vanished numerically when the type-II scheme is performed. This indicates that either the zero-mode contributions vanish or they have been included exactly into the valence contributions via the additive ``$M\to M_0$'' replacement. This conclusion is not rigorous but just a verification through a few of quantities. Therefore, it is meaningful to test the effectiveness of the ``$M\to M_0$'' replacement in more cases, for example, the EMFFs mentioned above. Moreover, if the zero-mode contributions in EMFFs are well dealt with, then the angular condition should also be well established. Consequently, the ambiguous from different prescriptions of EMFFs will be removed, {\it i.e.}, they will consistently give a set of numerical results. These are aims of this work.

This remaining parts of the paper are organized as follows. In section 2, we review briefly the EMFFs of the spin-1 particles and present four kinds of prescriptions for the form factors. In section 3, the zero-mode contributions are discussed in detail, and then the angular condition is investigated. Numerical results and discussions of the $\rho$-meson EMFFs $G_{C,M,Q}(Q^2)$ are presented in section 4. Section 5 is reserved for a summary.

\section{ELECTROMAGNETIC FORM FACTORS}\label{sec:2}
\label{sec:2}

For self-consistency, we first give a brief review on the EMFFs of spin-1 particles. Generally, the matrix elements of the one-body vector current $V^\mu=q''\r^\mu q'$ between two spin-1 particles can be divided into six independent Lorentz-invariant form factors~\cite{Jaus:2002sv}. For the case of electromagnetic transition, there are just three form factors $F_{1,2,3}(q^2)$ survived as follows \cite{Grach:1983hd,Carbonell:1998rj}:
\begin{eqnarray}
S^\mu_{h''h'}&=& \la M(p'')|V^\mu|M(p')\ra \nonumber\\
             &=& -\epsilon^*_{h''}\cdot \epsilon_{h'}P^\mu F_1(q^2) \nonumber\\
             & & +(\epsilon_{h'}^\mu P\cdot\epsilon_{h''}^*+\epsilon_{h''}^{*\mu} P\cdot\epsilon_{h'})F_2(q^2) \nonumber\\
             & & +\frac{(P\cdot\epsilon_{h'})(P\cdot\epsilon_{h''})}{2M^2}P^\mu F_3(q^2),
             \label{eq:amp1}
\end{eqnarray}
where $\epsilon^{(*)}$ and $h'~(h'')=0,\pm1$ denote the polarization vector and helicity of the initial (final) state, respectively. $P^\mu=p'^\mu+p''^\mu$ and $q^\mu=p'^\mu-p''^\mu$, where $p'~ (p'')$ is the 4-momentum of initial (final) state. For spin-1 particles, the condition of $p\cdot\epsilon=0$ is always hold. It should be kept in mind that the calculation of this work is performed with ``good'' plus component ($\mu=+$) in Drell-Yan frame. By setting $q^+=0$, $q^2=-q_\perp^2\equiv-Q^2$ denotes the space-like momentum transfer. The physical charge, magnetic and quadrupole form factors $G_{C,M,Q}(q^2)$ can be determined by $F_i(q^2)$ via the following relations~\cite{Arnold:1979cg}:
\begin{align}
G_C(q^2)&=F_1(q^2)+\frac{2}{3}\eta G_Q(q^2),~~\nonumber\\
G_M(q^2)&=-F_2(q^2),~~\nonumber\\
G_Q(q^2)&=F_1(q^2)+F_2(q^2)+(1+\eta)F_3(q^2),
\label{eq:GCMQ}
\end{align}
where the kinematic factor $\eta=\frac{Q^2}{4M^2}$, and as will be shown later, the form factors $G_{C,M,Q}(q^2)$ can be expressed through matrix elements $S^{+}_{h''h'}$.

There are nine matrix elements $S^+_{h''h'}$, which correspond to nine modes of transition between the initial and final polarization states. Not all of the matrix elements $S^+_{h''h'}$ are independent, which are constrained by (i) the time-reversal invariance, (ii) the reflection invariance in the $(\hat{x},\hat{y})$ plane, and (iii) the rotation invariance about $\hat{z}$~\cite{Bakker:2002mt}. As a result, there are only four independent matrix elements, which are usually chosen as $S^+_{00}$, $S^+_{10}$, $S^+_{1-1}$ and $S^+_{11}$. Actually, those four $S^+_{h''h'}$ are not independent if the rotational covariance is completely satisfied but not just hold for the rotation about $\hat{z}$. If the rotational covariance is completely satisfied, they should be restrained by an additive relation equation, which is the so-called angular condition~\cite{Grach:1983hd}:
\begin{align}\label{eq:ac}
\Delta(Q^2)&\equiv (1+2\eta)S^+_{11}-\sqrt{8\eta}S^+_{10}+S^+_{1-1}-S^+_{00}=0.
\end{align}
In other words, the number of independent matrix elements changes down to three when taking the angular condition into consideration.

For the light-front approaches, the Hamiltonian $P^-$ is usually chose as $P^0-P\cdot\hat{n}$, where $\hat{n}$ is the spatial unit vector. Practically, $\hat{n}$ is usually set along with the $\hat{z}$ axis and leads to $P^-=P^0-P^3$. This choice will also be taken in this work~\footnote{In deed, we take the BL convention and there is an additional ``$1/2$'' factor in $P^-=\frac{1}{2}(P^0-P^3)$. The result is independent to the conventions for calculating.}. A variety of contexts for studies of rotational covariance have been performed in Refs.\cite{Chung:1988mu, Grach:1983hd, Carbonell:1998rj, Keister:1993mg, Cardarelli:1994yq}. The point is that the approximated one-body current is free of interaction, then the matrix elements of one-body current cannot satisfy the complete rotational covariance or angular condition in principle~\cite{Jaus:2002sv}~\footnote{It is an abstruse task to recovered the complete rotational covariance via an in-depth way, and in the paper, we will adopt the same approximation for rotational covariance.}. And because the rotation covariance cannot not completely satisfied, e.g. the condition (iii) is adopted for discussion, we will have $\Delta(q^2)\neq 0$. This underlies why the form factors $G_{C,M,Q}(q^2)$ vary from each others in the works of Refs.\cite{Grach:1983hd, Chung:1988my, Brodsky:1992px, Frankfurt:1993ut}.

For the sake of conciseness, we list four versions $G_{C,M,Q}(q^2)$ that is derived under different linear combinations of $S^+_{h''h'}$. The first one is suggested by Grach and Kondratyuk~\cite{Grach:1983hd}
\begin{align}\label{eq:GGK}
G_C^{\rm GK}(q^2)&=\frac{1}{3}[(3-2\eta)S^+_{11}+2\sqrt{2\eta}S^+_{10}+S^+_{1-1}],\nonumber\\
G_M^{\rm GK}(q^2)&=2[S^+_{11}-\frac{1}{\sqrt{2\eta}}S^+_{10}],\nonumber\\
G_Q^{\rm GK}(q^2)&=\frac{2\sqrt{2}}{3}[-\eta S^+_{11}+\sqrt{2\eta}S^+_{10}-S^+_{1-1}],
\end{align}
where the matrix element $S^+_{00}$, which contains the zero-mode contributions, has no contribution to those form factors. The second one is suggested by Chung, Coester, Keisbter and Polyzou~\cite{Chung:1988my}
\begin{align}\label{eq:GCCKP}
G_C^{\rm CCKP}(q^2)&=\frac{1}{3(1+\eta)}[(\frac{3}{2}-\eta)(S^+_{11}+S^+_{00})+5\sqrt{2\eta}S^+_{10}\nonumber\\
&+(2\eta-\frac{1}{2})S^+_{1-1}],\nonumber\\
G_M^{\rm CCKP}(q^2)&=\frac{1}{(1+\eta)}[S^+_{11}+S^+_{00}-\frac{2(1-\eta)}{\sqrt{2\eta}}
S^+_{10}-S^+_{1-1}],\nonumber\\
G_Q^{\rm CCKP}(q^2)&=\frac{\sqrt{2}}{3(1+\eta)}[-\eta(S^+_{11}+S^+_{00})+
2\sqrt{2\eta}S^+_{10}\nonumber\\
&-(\eta+2)S^+_{1-1}].
\end{align}
The third one is suggested by Brodsky and Hiller~\cite{Brodsky:1992px}
\begin{align}\label{eq:GBH}
G_C^{\rm BH}(q^2)&=\frac{1}{3(1+\eta)}[(3-2\eta)S^+_{00}+
8\sqrt{2\eta}S^+_{10}\nonumber\\
&+2(2\eta-1)S^+_{1-1}],\nonumber\\
G_M^{\rm BH}(q^2)&=\frac{2}{(1+2\eta)}[S^+_{00}+\frac{2\eta-1}{\sqrt{2\eta}}
S^+_{10}-S^+_{1-1}],\nonumber\\
G_Q^{\rm BH}(q^2)&=\frac{2\sqrt{2}}{3(1+2\eta)}[-\eta S^+_{00}+
\sqrt{2\eta}S^+_{10}-(\eta+1)S^+_{1-1}],
\end{align}
where $S^+_{00}$ is treated as the dominant one at large $Q^2$ by perturbative QCD analyzes, while $S^+_{11}$ has no contribution to those form factors. And the fourth one is suggested by Frankfurt, Frederico and Strikman~\cite{Frankfurt:1993ut}
\begin{align}\label{eq:GFFS}
G_C^{\rm FFS}(q^2)&=\frac{1}{3(1+\eta)}[(3+2\eta)S^+_{11}\nonumber\\
&+2\sqrt{2\eta}S^+_{10}+(2\eta+1)S^+_{1-1}-\eta S^+_{00}],\nonumber\\
G_M^{\rm FFS}(q^2)&=G_M^{\rm CCKP}(q^2),\nonumber\\
G_Q^{\rm FFS}(q^2)&=G_Q^{\rm CCKP}(q^2).
\end{align}

In the light-front quark model, the full result is generally composed by the valence part and the zero-mode contributions, i.e. $\tilde{Q}_{\text{full}}=\tilde{Q}_{\text{val.}}+\tilde{Q}_{\text{z.m.}}$. Refs.\cite{Bakker:2002mt, Choi:2004ww, deMelo:1997cb, Jaus:2002sv} show that only the helicity zero-to-zero matrix element $S^+_{00}$ receives the zero-mode contribution and the others just contain the valence parts. In addition, the authors there claim that if the zero-mode contribution is include in $S_{00}^+$, the zero-mode contribution should be closely associated with the angular conditions and $\Delta(q^2)=0$. So in the following, we give a discussion on the zero-mode contribution and the angular condition.

\section{ZERO-MODE CONTRIBUTION AND ANGULAR CONDITION}

In the standard light-front (SLF) quark model, the constituent quark and antiquark in a bound state are required to satisfy their respective on-shell conditions, with physical quantities calculated directly in three-dimensional light-front momentum space. Typically, the ``good" component ``$\mu=+$" of the current matrix elements is selected to avoid the zero-mode contributions. It is particularly noteworthy that the matrix elements obtained in the SLF quark model lose Lorentz covariance, and the model itself cannot independently determine the zero-mode contributions. Compared with the SLF quark model, the covariant light-front (CLF) quark model exhibits two distinctive features: First, it provides a systematic approach to study the zero-mode effects; Second, by employing the manifestly covariant Bethe-Salpeter (BS) method, it's predictions satisfy Lorentz covariance. It must be emphasized that the CLF quark model introduces an effective vertex $\chi_M$ between quarks and bound states, whose specific form needs to be determined by matching with the results from the SLF quark model. Subsequently, to address the self-consistency issue of vector meson decay constants in LFQM, Choi and Ji proposed the following two replacement schemes~\cite{Choi:2013mda}:
\begin{eqnarray}
\chi_M &\rightarrow& h_M/\hat{N}=\frac{1}{\sqrt{2N_c}}\sqrt{\frac{\bar{x}}{x}}\frac{\psi_s}{\hat{M}_0}, \nonumber\\
D_{\rm con} &=& M+m_1+m_2 \nonumber\\
            &\rightarrow& D_{\rm LF}=M_0+m_1+m_2 \quad (\text{Type-I}) \label{eq:typeI}
\end{eqnarray}
and
\begin{eqnarray}
\chi_M &\rightarrow& h_M/\hat{N},~~M\rightarrow M_0, \quad (\text{Type-II}) \label{eq:typeI}
\end{eqnarray}
where $\bar{x}=1-x$ and $x$, both of them are the momentum fractions carried by the constituent quarks, while $\psi_{s}$ corresponds to the $s$-wave wavefunction of the vector meson. $M$ and $M_0$ are physical mass and invariant mass of the vector meson, respectively, e.g. $M^2_0=(m^2_1+k^2_\perp)/{x}+(m^2_2+k^2_\perp)/{\bar{x}}$ and $\hat{M}^2_0=M^2_0-(m_1-m_2)^2$. The constant factor $D_{\rm con}$ and the light-front factor $D_{\rm LF}$ emerge from the vertex structure of the meson's spin wavefunction. The distinction between Type-I and Type-II schemes lies in the additive ``$M\to M_0$" replacement for all the mesons' masses in type-II formulas. Actually, the effective vertex function $\chi_M$ just appears in the CLF quark model, while it is absent in the SLF quark model. So the replacement for $\chi_M$ is redundant in SLF quark model. Therefore, the SLF quark model does not have the same two substitution options, it can only perform the ``$M\to M_0$" replacement.

In doing numerical analyses, $\rho$ meson is taken as an example to deploy the explanation, whose input parameters are $m_q=0.25\pm0.04$ GeV and $\beta_{q\bar q}=0.3124\pm0.0060$ GeV that are derived by fitting the experiment results of the decay constants $f_\pi$ and $f_\rho$~\cite{Chang:2018zjq}.

\subsection{ZERO-MODE CONTRIBUTION}
\label{sec:3}
Refs.\cite{Chang:2018zjq, Chang:2019mmh} have analyzed the problems of self-consistency and the covariance of LFQM by studying the mesonic decay constants and the weak transition form factors. It has been found that the self-consistency problem is closely related with the covariance problem, both of them are associate with the zero-mode contribution; and by using the type-II scheme, two problems can be naturally solved. This is caused by the fact that even though the zero-mode contribution will be in the formulas, but their numerical magnitude becomes zero when the type-II replacement has been done. The validity of type-II scheme has also been tested through the $V'\to V''$ ($V$ denotes the vector meson) weak transition processes~\cite{Chang:2019obq}. It is worth investigating whether the type-II scheme can also contribute to solving the self-consistency problem in $\rho$-meson EMFFs.

The complete results of the CLF quark model can be expressed as the sum of valence contributions and zero-mode contributions, $S^{\text{CLF}}=S^{\text{val}}+S^{\text{z.m.}}$. Detailed analysis reveals distinct behaviors for different matrix elements: For $S_{11}^{+}$, $S_{10}^{+}$, and $S_{1-1}^{+}$, the valence contributions are found to be identical to the full results, indicating the absence of zero-mode contributions, and under the type-I scheme, these matrix elements show exact formal agreement between the SLF and CLF quark models. This has also been confirmed in previous studies~\cite{Brodsky:1992px, Chang:2018zjq, Chang:2019mmh, Chang:2019obq}.

\begin{figure}[htb]
\centering
\includegraphics[width=0.45\textwidth]{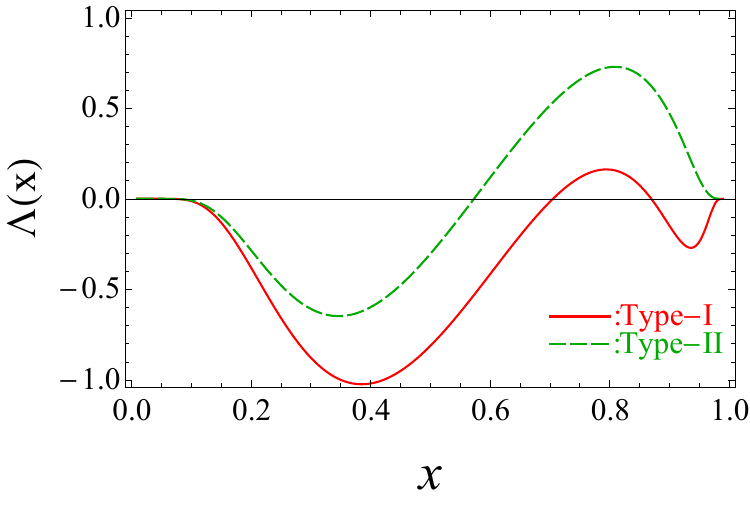}
\caption{The solid and dashed lines are for $\Lambda(x)$ versus $x$ under the type-I and the type-II schemes, respectively. $\Lambda(x)$ shows the differences between $S_{00}^{+\text{CLF}}$ and $S_{00}^{+\text{SLF}}$.}
\label{fig:Szz}
\end{figure}

However, the situation differs significantly for the matrix element $S_{00}^{+}$, where under type-I scheme we observe $S_{00}^{+\text{CLF}}\neq S_{00}^{+\text{val}}$, demonstrating the involvement of zero-mode contributions in this case. Remarkably, when employing the type-II scheme, two important features emerge: first, $S_{00}^{+\text{CLF}}$ and $S_{00}^{+\text{val}}$ yield identical numerical values across all momentum transfers $Q^2$; second, the relation $S_{00}^{+\text{CLF}}=S_{00}^{+\text{val}}=S_{00}^{+\text{SLF}(M\rightarrow M_0)}$ is satisfied. These findings collectively demonstrate that the type-II scheme provides an effective approach for addressing computational discrepancies arising from zero-mode contributions in the framework of LFQM. And to show their differences, we define a function $\Lambda(x)$:
\begin{align}\label{eq:Szz}
\Lambda(x)\equiv\frac{d}{dx}S_{00}^{+\text{CLF}}-\frac{d}{dx}S_{00}^{+\text{SLF}(M\rightarrow M_0)}.
\end{align}
 The value of $\int_0^1 \Lambda(x) \, dx$ corresponds to the zero-mode contribution, since the SLF result always equals the valence contribution under the $M \to M_0$ mass replacement scheme. We present $\Lambda(x)$ versus $x$ in Fig.~\ref{fig:Szz}, where the solid and dashed lines are results of the type-I and the type-II schemes, respectively. It indicates $\int_{0}^{1}\Lambda(x)dx\neq0$ for the type-I scheme, which vanishes exactly for the type-II scheme. This confirms the previous findings of Refs.\cite{Chang:2018zjq, Chang:2019mmh, Chang:2019obq} and reveals the feasibility of type-II scheme. It means that either there is no zero-mode contribution or it has been included into the valence part through the additive replacement of ``$M\to M_0$".

\subsection{ANGULAR CONDITION}

\begin{figure}[htb]
\centering
\includegraphics[width=0.45\textwidth]{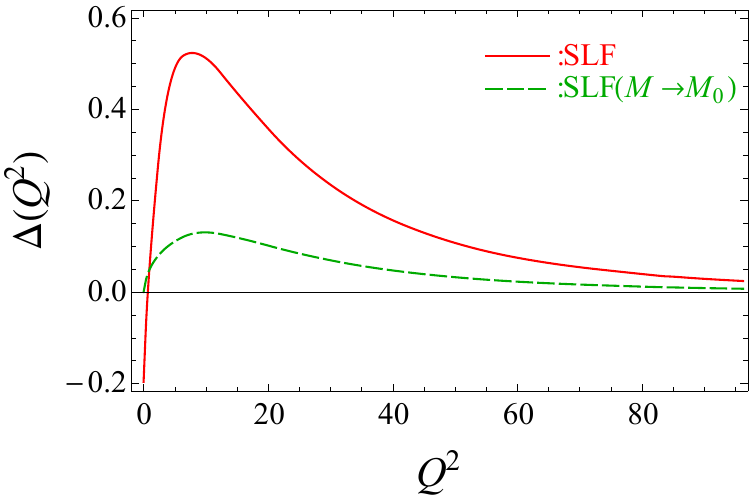}
\caption{The $\rho$-meson angular condition $\Delta(Q^2)$ versus $Q^2$. The dashed and solid lines are for $\Delta(Q^2)$ under the SLF and the SLF($M\to M_0$), respectively.}
\label{figAC}
\end{figure}

We present the $\rho$-meson angular condition $\Delta(Q^2)$ versus $Q^2$ in Fig.~\ref{figAC}, where the dashed and solid lines are for $\Delta(Q^2)$ under the case of SLF or the SLF($M\to M_0$). For the case of SLF, $\Delta(Q^2)$ becomes negative in small $Q^2$ region ($Q^2\to 0$), leading to visibly violation of angular condition under the SLF quark model. In Ref.\cite{Choi:2004ww}, the authors claimed that the zero-mode contribution in $S_{00}^+$ should be vanished in the zero-binding limit ($M\to M_0$), and then the angular condition is hold at $Q^2=0$. We confirm their observation. More explicitly, we have $\Delta(0)=0$ for the case of SLF($M\to M_0$), indicating the angular condition is strictly satisfied at the zero-momentum transfer. It is noted that the above mentioned four versions of EMFFs (\ref{eq:GGK}, \ref{eq:GCCKP}, \ref{eq:GBH}, \ref{eq:GFFS}) exactly give the same values at $Q^2=0$ for the case of type-II scheme. They also yield consistent results in the low regions where $Q^2\lesssim0.5~\text{GeV}^2$. In contrast, the results of SLF are inconsistent in these regions, as can be seen below.

For the case of SLF, Fig.~\ref{figAC} shows that in moderate $Q^2$ region, $\Delta(Q^2)$ will first increase and the decrease with the increment of $Q^2$, whose maximum value $\simeq 0.53$ is achieved for $Q^2\simeq 10\text{GeV}^2$ and the angular condition is significantly damaged, which will lead to different results by using different prescriptions for the EMFFs. For the case of SLF($M\to M_0$), $\Delta(Q^2)$ behaves similarly and its maximum value $\simeq 0.13$ is achieved for $Q^2\simeq 10\text{GeV}^2$. Thus strictly speaking, the angular condition is not satisfied in moderate $Q^2$ region even for the case of SLF($M\to M_0$). Nevertheless, the violation of angular condition can be remarkably softened for the case of SLF($M\to M_0$), which can be safely neglected for the EMFFs, as their differences under different versions of EMFFs are small. For both cases, the asymptotic value of $\Delta(Q^2)$ is zero when $Q^2\to\infty$. The asymptotic behavior of $\Delta(Q^2)$ at large $Q^2$ transfer is similar to the result of Ref.\cite{Choi:2004ww}, which is however quite different from the results of Refs.~\cite{Grach:1983hd, Keister:1993mg}, in which the angular condition is violated badly at high $Q^2$ region. One may observe $\Delta(Q^2\simeq0.7\text{GeV}^2)=0$, which indicates that for the case of SLF the angular condition can be achieved at this point.

\begin{figure}[htb]
\centering
\includegraphics[width=0.45\textwidth]{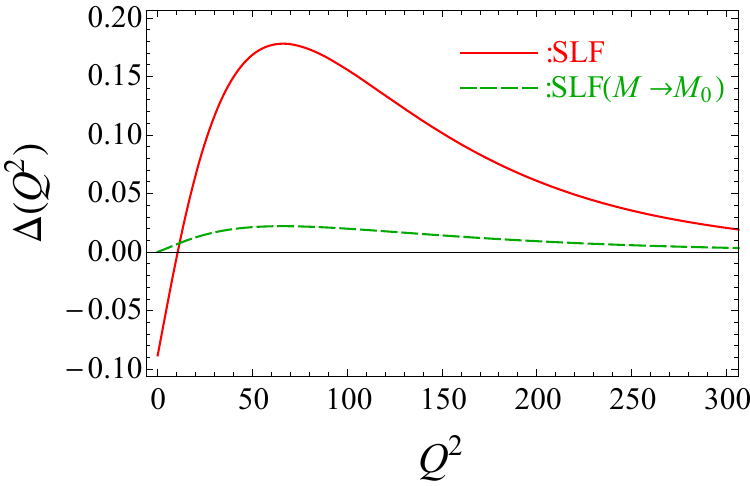}
\caption{The $\Upsilon(1S)$-meson angular condition $\Delta(Q^2)$ versus $Q^2$. The dashed and solid lines are for $\Delta(Q^2)$ under the SLF and the SLF($M\to M_0$), respectively.}
\label{fig:ACbb}
\end{figure}

Earlier studies on the deuteron indicate that $\Delta(Q^2)$ can be relatively small~\cite{Grach:1983hd, Chung:1988my, Keister:1993mg}, where the important feature of the deuteron is the characteristic nucleon momentum is very small compared to its nucleon mass. We analogously carry out the calculation about heavy meson $\Upsilon(1S)$, in which the mass of the constituent $b (\bar b)$ is large than the characteristic momentum. It is expected that the value of $\Delta(Q^2)$ should be quite smaller than the case of $\rho$ meson above. As shown in Fig.~\ref{fig:ACbb}, the violation effects do dramatically decline for both cases of SLF ($\Delta(Q^2)<0.2$) and SLF($M\to M_0$) ($\Delta(Q^2)<0.02$). It is worth mentioning that the mass $M$ and invariant mass $M_0$ are equivalent in the heavy quark limit.

In general, the angular condition is unfulfillable and thereupon the four kinds of prescriptions cannot give a consistent result in SLF. After applying the transformation ($M\to M_0$), the violation effect of angular condition shall be remarkably suppressed and becomes negligible, even through it does not vanish entirely. The SLF($M\to M_0$) greatly improves the situation of the angular condition and therefore different versions of EMFFs will give consistent values.

Our result agrees with the previous observation of Refs.\cite{Grach:1983hd, Bakker:2002mt} that the matrix element $S_{00}^+$ is the only one that contains the zero-mode contribution and the GK's version do be immune to the zero-mode contribution by deliberately avoiding $S_{00}^+$ in their formulas. From this perspective, it is definitely an advantage among the four prescriptions tested. However, it seems like an expedient to avoiding zero-mode contribution by the assumption of angular condition.

\section{NUMERICAL RESULTS}

\subsection{The EMFFs $G_{C,M,Q}(Q^2)$}

\begin{figure}[htb]
\centering
\includegraphics[width=0.45\textwidth]{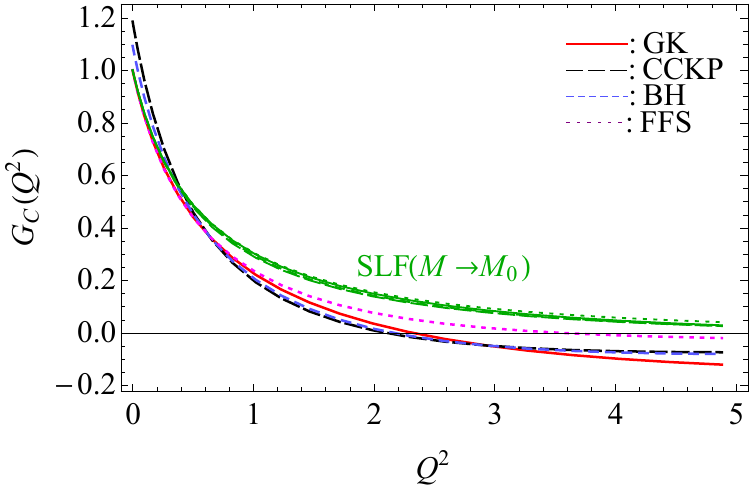}
\caption{The charge form factor $G_C(Q^2)$ versus $Q^2$ for $\rho$ meson. Different lines are for four versions of EMFFs under the prescriptions of GK, CCPK, BH and FFS, respectively. The red-solid, black-dashed, blue-short-dashed and purple-dotted lines are for case of SLF. The same line types but with green color are for case of SLF($M\to M_0$).}
\label{figGC}
\end{figure}

\begin{figure}[htb]
\centering
\includegraphics[width=0.45\textwidth]{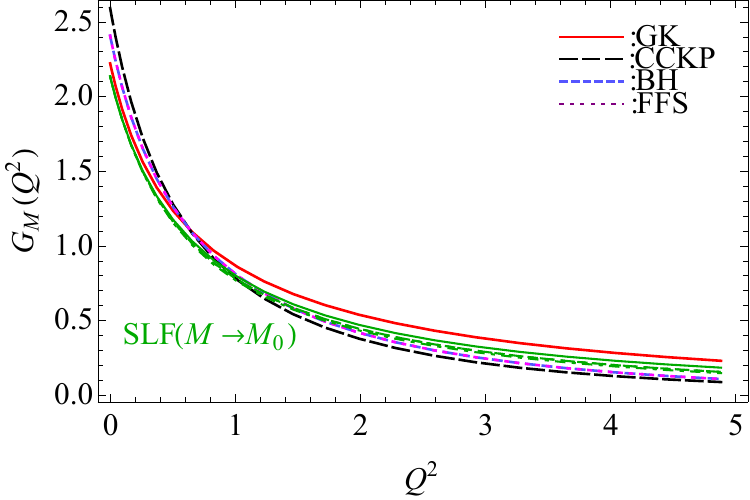}
\caption{The magnetic form factor $G_M(Q^2)$ versus $Q^2$ for $\rho$ meson. Different lines are for four versions of EMFFs under the prescriptions of GK, CCPK, BH and FFS, respectively. The red-solid, black-dashed, blue-short-dashed and purple-dotted lines are for case of SLF. The same line types but with green color are for case of SLF($M\to M_0$).}
\label{figGM}
\end{figure}

\begin{figure}[htb]
\centering
\includegraphics[width=0.45\textwidth]{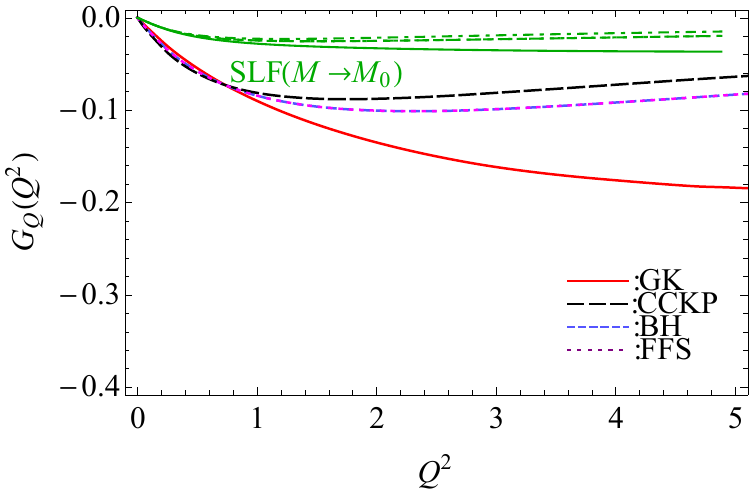}
\caption{The quadrupole form factor $G_Q(Q^2)$ versus $Q^2$ for $\rho$ meson. Different lines are for four versions of EMFFs under the prescriptions of GK, CCPK, BH and FFS, respectively. The red-solid, black-dashed, blue-short-dashed and purple-dotted lines are for case of SLF. The same line types but with green color are for case of SLF($M\to M_0$).}
\label{figGQ}
\end{figure}

Based on the above discussions, we calculate the $\rho$ meson EMFFs $G_{C,M,Q}(Q^2)$ in space-like momentum transfer under four prescriptions GK, CCPK, BH and FFS. Our results for the SLF and the SLF($M\to M_0$) are presented in Figs.~\ref{figGC}, \ref{figGM} and \ref{figGQ}. Fig.~\ref{figGC} shows that different prescriptions will bring discrepant values for charge form factor $G_C(Q^2)$ under the SLF, which is just the consequence of $\Delta(Q^2)\neq 0$. Around $Q^2\simeq0.7\text{GeV}^2$, the four lines under the SLF intersect with each other, because the angular condition is satisfied at that special point as shown by Fig.\ref{figAC}. With the increment of $Q^2$, the discrepancies among the four prescriptions become noticeable expect that of CCKP and BH. Generally, the violation of angular condition is not negligible. One mentionable thing is that $G_C(Q^2)$ under the SLF goes to zero at some specific point of $Q^2$, such as $Q^2\simeq3.5\text{GeV}^2$ for FFS and $Q^2\simeq2.2\text{GeV}^2$ for the other prescriptions. This agrees with previous observation of Ref.\cite{Arnold:1979cg} at large $Q^2$, $G_C(Q^2)$ becomes negative. While for the case of SLF($M\to M_0$), the results under four prescriptions wholly overlap with each other at low momentum transfer $Q^2<0.5\text{GeV}^2$, which is the consequence of $\Delta(Q^2)\simeq0$. And the deviations of their values at intermediate and large momentum transfers are quite small which are limited in the uncertainty from inputs and can be safely neglected. Phenomenologically, it is acceptable that the four prescriptions give the same predictions. Thus the SLF($M\to M_0$) improves greatly the consistence of different prescriptions.

Fig.~\ref{figGM} exhibits the $Q^2$-dependence of magnetic form factor $G_M(Q^2)$ from the four prescriptions under the SLF and SLF($M\to M_0$), respectively. Similar to the case of $G_C(Q^2)$, different prescriptions give obviously different results for the case of SLF, especially for the GK's. By applying the ``$M\to M_0$" replacement, similar to the case of $G_C(Q^2)$, all the curves of $G_M(Q^2)$ for the four prescriptions are consistent with each other.

The quadrupole form factor $G_Q(Q^2)$ is shown in Fig.~\ref{figGQ}, where the diversity of vary prescriptions seems more pronounced for both the SLF and the SLF($M\to M_0$). Furthermore, the difference between results of the SLF and the SLF($M\to M_0$) intensifies compared with that of $G_C(Q^2)$ and $G_M(Q^2)$. It is need to say that the violation effect of angular condition is same for all the form factors $G_C(Q^2)$, $G_M(Q^2)$ and $G_Q(Q^2)$. While the more remarkable influence for $G_Q(Q^2)$ is just the result of its value is relatively small. The value of $G_Q(Q^2)$ reflects the relativistic effects~\cite{deMelo:1997cb}, which is the reason that $G_Q(0)=0$ and its module increase by the bigger $Q^2$. Comparing the different types of lines, one can find that the values of $G_Q(Q^2)$ from the four prescriptions  under the case of SLF($M\to M_0$) are close to each other, indicating that the relativistic effects in $\rho$ meson are small in this case.

As a summary, by applying the ``$M\to M_0$" replacement, it has been discovered that the discrepancies among the four prescriptions can be significantly reduced, yielding a consistent result for all the prescriptions.

\subsection{The mean square radius $\langle r^2 \rangle$, magnetic moment $\mu$ and quadrupole moment $\bar Q$}

\begin{table}[htb]
\centering
\begin{tabular}{|c|c|c|c|c|c|c|}
\hline
    &{GK} &{CCKP} &{BH} &{FFS} &{SLF($M\to M_0$)}&{Previous}\\\hline
  $\langle r^2 \rangle$&0.48&---&---&0.49&0.47&$0.35-0.40$\\\hline
  $\bar Q$ &$0.022$&$0.026$&$0.030$&$0.026$&$0.010$&$0.024-0.058$\\\hline
  $\mu$    &2.23 &2.36 &2.48 &2.36 &2.13&$2.14-2.48$\\\hline
\end{tabular}
\caption{The charge radius $\langle r^2 \rangle$, magnetic momentum $\mu$ and quadrupole momentum $\bar Q$ with and without ``$M\to M_0$" replacement in the SLF quark model, where previous results~\cite{Chung:1988mu, Grach:1983hd, Carbonell:1998rj, Keister:1993mg, Cardarelli:1994yq} have also been listed as a comparison.}
\label{tab:gz}
\end{table}

The charge radius $\langle r^2 \rangle$, magnetic moment $\mu$ and quadrupole moment $\bar Q$ can be defined as~\cite{Keister:1993mg}
\begin{eqnarray}\label{eq:gz}
\langle r^2\rangle &=& \lim_{Q^2\to0}\frac{6[1-G_C(Q^2)]}{Q^2},~~~~\nonumber\\
\mu &=& \lim_{Q^2\to0}G_M(Q^2),~~~~\nonumber\\
\bar Q &=& \lim_{Q^2\to0}\frac{3\sqrt{2}G_Q(Q^2)}{Q^2}.
\end{eqnarray}
We also calculate these quantities with and without ``$M\to M_0$" replacement in SLF and present them in table~\ref{tab:gz}. The charge radius $\langle r^2 \rangle$ reflects the slope of physical form factor $G_C(Q^2)$ at $Q^2=0$ and it should be careful to calculate this quantity as the possible singular behaviors of some cases. From the first row of table ~\ref{tab:gz}, it is found that the results of $\langle r^2 \rangle$ for GK's and FFS's versions under SLF are consistent with each other, and slightly larger than the result of SLF($M\to M_0$). Compared with the previous works, where $\langle r^2 \rangle\sim(0.35-0.40)~\text{fm}^2$, our number increases by $25\%$, which indicates the size of $\rho$ meson is about $0.7~\text{fm}$ in this respect. The magnetic momentum $\mu$ equals to 2 in the limit that the spin-1 particle becomes a point state and $\mu-2$ defines the anomalous momentum, which reflects the dynamical contributions due to internal structure~\cite{Jaus:2002sv}. For instance, the nonrelativistic SU(6) model predicts $\mu=2$, where all interaction effects are neglected. It is noticeable that $\mu=2.13$ under SLF($M\to M_0$) is smaller than that of SLF $\mu\sim2.23-2.48$, which implies the ``$M\to M_0$" replacement weaken the relativistic effect or the interaction in bound state. Similarly, the quadrupole momentum $\bar Q$ with SLF($M\to M_0$) is smaller than the results with SLF, which is also leaded by the decrescendo of relativistic effect.

\section{CONCLUSIONS}

In summary, the EMFFs of $\rho$ meson have been studied in the light-front quark model. It is confirmed that only $S_{00}^+$ contains the zero-mode contribution. Numerical analyses on the matrix element $S_{00}^+$ shows the valence and full results are exactly equivalent. This implies that either the zero-mode contribution vanishes or they have been included into the valence part via the type-II and the ``$M\to M_{0}$" replacements in the CLF and SLF quark models, respectively. It is found the ``$M\to M_0$" replacement improves the angular condition significantly and different prescriptions for the physical EMFFs can give a consistent result. Numerical analysis indicates that the ``$M\to M_0$" replacement weaken the relativistic effects in bound stats. The residuary minor violation of angular condition needs some other explanations.

\section*{Acknowledgements}

We thank Qin Chang for many helpful discussions. The work is supported by the National Natural Science Foundation of China (Grant No.11875122, No.12175025 and No.12147102) and Tongling University Talent Program (Grant No.R23100).

\end{document}